# Initial recommendations for performing, benchmarking, and reporting single-cell proteomics experiments

Laurent Gatto, Ruedi Aebersold, Juergen Cox, Vadim Demichev, Jason Derks, Edward Emmott, Alexander M. Franks, Alexander R. Ivanov, Ryan T. Kelly, Luke Khoury, Andrew Leduc, Michael J. MacCoss, Peter Nemes, David H. Perlman, Aleksandra A. Petelski, Christopher M. Rose, Erwin M. Schoof, Jennifer Van Eyk, Christophe Vanderaa, John R. Yates III, and Nikolai Slavov

Supporting website: https://single-cell.net/guidelines
Correspondence: nslavov@northeastern.edu

**Analyzing proteins from single cells by tandem mass spectrometry (MS) has become technically feasible. While such analysis has the potential to accurately quantify thousands of proteins across thousands of single cells, the accuracy and reproducibility of the results may be undermined by numerous factors affecting experimental design, sample preparation, data acquisition, and data analysis. Broadly accepted community guidelines and standardized metrics will enhance rigor, data quality, and alignment between laboratories. Here we propose best practices, quality controls, and data reporting recommendations to assist in the broad adoption of reliable quantitative workflows for single-cell proteomics.**

New approaches and technologies for experimental design, sample preparation, data acquisition, and data analysis have enabled the measurement of several thousand proteins in small subpopulations of cells and even in single mammalian cells[1–11]. These developments open exciting new opportunities for biomedical research[12], as illustrated in Fig. 1. In some systems, subpopulations of molecularly and functionally similar cells can be isolated and analyzed in bulk, which allows for deeper proteome coverage. Other systems, however, do not allow for such isolation due to continuous (rather than discrete) phenotypic states, or due to unknown cell states or markers[13,14]. Such systems require single-cell analysis; it is particularly needed for discovering new cell types[15] and for investigating continuous gradients of cell states, which has already benefited from single-cell MS proteomics[6,16–18]. Furthermore, when a large number of single cells are analyzed, the joint distributions of protein abundances enable new types of data-driven analysis (Fig. 1) that may support inferences with minimal assumptions[12,19].

Despite these promising prospects, single-cell MS is sensitive to experimental and computational artifacts that may lead to failures, misinterpretation, or substantial biases that can compromise data quality and reproducibility, especially as the methodologies become widely deployed. To minimize biases and to maximize quantitative accuracy and reproducibility of single-cell proteomics, we propose initial guidelines for optimization, validation, and reporting of single-cell proteomic workflows and results.

The tandem MS methods for single-cell bottom-up proteomics span a range of techniques[13], including multiplexed and label-free methods, both of which can be performed by data-dependent acquisition (DDA)[1,20] and data-independent acquisition (DIA)[7,10]. The initial recommendations presented here are relevant to all of these methods, and we will note any exceptions. Our initial recommendations intend to stimulate further community-wide discussions that mature into robust, widely adopted practices. Imaging and top-down MS methods are also advancing and reaching single-cell resolution[21], although they differ significantly from MS-based bottom-up proteomics methods and are outside the scope of this white paper. Our recommendations are topically grouped into "experimental design", "data evaluation and interpretation", and "reporting".

# Experimental design

Best practices for single-cell MS proteomics can effectively build on established practices for bulk analysis[22,23]. Common best practices include staggering biological treatments, sample processing, and analytical batches so that sources of biological and technical variation can be distinguished and accounted for during results interpretation. Similarly, randomization of biological and technical replicates and batches of reagents during sample processing (e.g., mass tags for barcoding) are recommended to minimize potential artifacts and to facilitate their diagnoses. We also recommend including appropriately diluted bulk samples as technical quality controls. The following specific issues are relevant for the design of single-cell proteomic measurements.

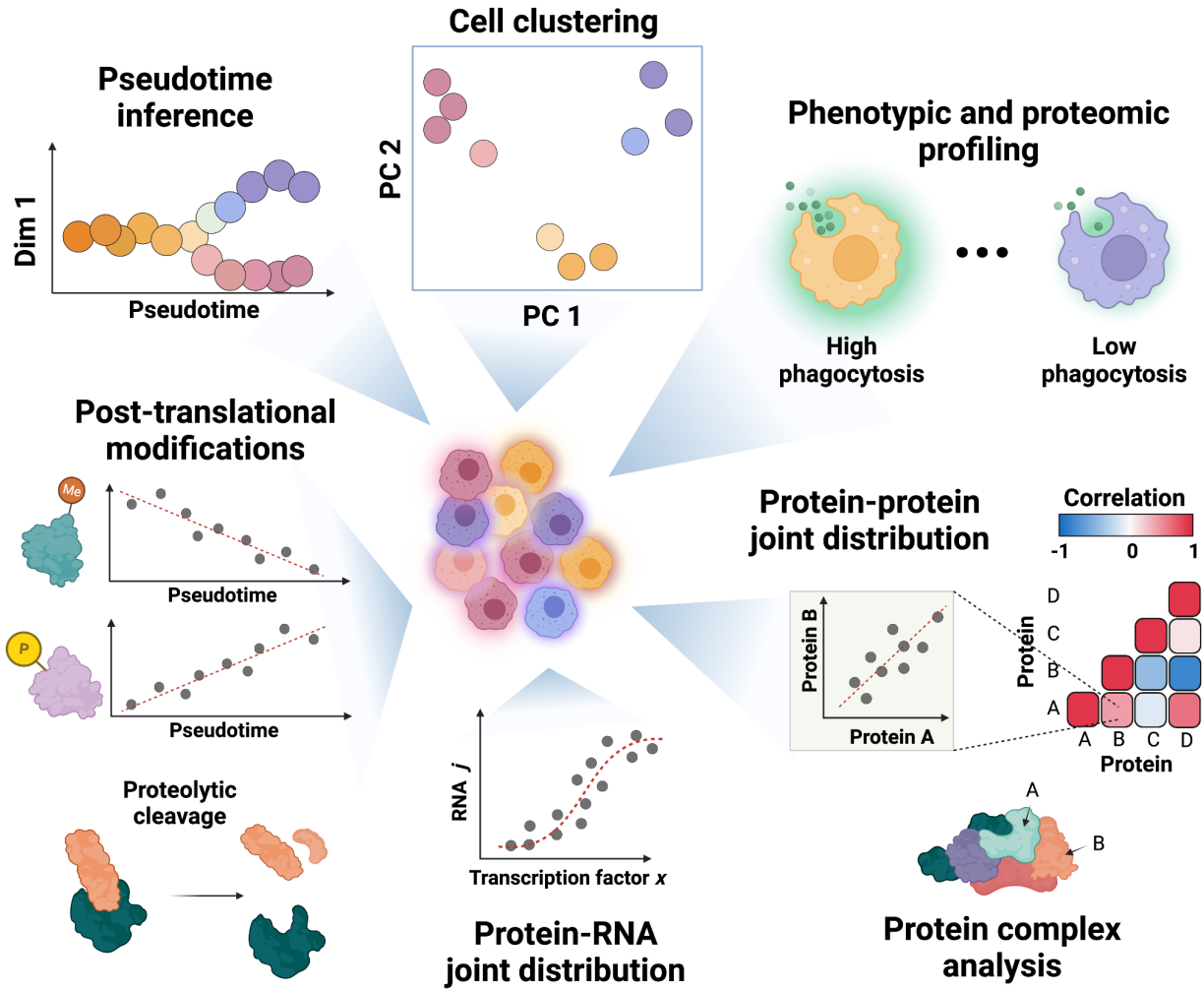

**Figure 1 | Emerging applications of single-cell proteomics by mass spectrometry.** Single-cell proteomic measurements can define cell-type and cell-state clusters[9], support pseudotime inference, link protein levels to functional phenotypes, such as phagocytic activity[18], quantify protein covariation and apply it to study protein complexes[1,6,19], analyze protein conformations[24], and quantify protein modifications, such as phosphorylation and proteolysis[5,6,18]. Furthermore, integrating protein and RNA measurements from the same biological systems (as in ref.[1,16]) allows inferring transcriptional and posttranslational regulation[1,16] investigating the covariation of transcription factors and downstream target transcripts[16].

**Single-cell isolation.** A primary goal of sample preparation should be to preserve the biological state of the cells with minimal perturbations. This can be challenging for tissues and for adherent cell cultures as cell isolation may require vigorous dissociation or detachment procedures. Extracting single cells from tissue samples in some cases may require enzymatic digestion of proteins, which may cleave the extracellular domains of surface proteins. Potential artifacts arising from these manipulations should be considered and may be minimized by using more gentle dissociation procedures, such as chelation of cations stabilizing extracellular protein interactions. Dissociated single cells should be thoroughly washed to minimize contamination of MS samples with reagents used for tissue dissociation.

While proteins are generally more stable than mRNAs[25], most good practices utilized for isolating cells for single-cell RNA-seq and flow cytometry[26], such as quick sample processing at low temperature (4°C), are appropriate for proteomics as well. The timing and other parameters of the cell isolation procedure may be impactful and so should be recorded in order that technical effects associated with sample isolation can be accounted for in downstream analysis. We recommend collecting as much phenotypic information as possible from cells prepared and isolated in the same manner, including cellular images and any relevant functional assays that can be performed. Such phenotypic data allow for orthogonal measures of cell state to be combined with mass spectrometry data and thus to strengthen biological interpretations. While isolating single cells of interest, we recommend also collecting bulk samples from the same cell population (if possible). Having such bulk samples will allow for the inclusion of positive controls and for benchmarking; these two topics will be discussed more in sections below.

Many studies have used FACS for isolating cells from a single-cell suspension[9,10,16,27]. FACS can perform very well, as indicated by the successful results of such studies. Yet, in the absence of high performing sorters and expert operators, it may be one of the least robust steps of the workflow[5]. Thus, verifying the ability to robustly isolate individual cells by FACS may save much time from troubleshooting downstream analysis steps. Studies have also isolated single cells by CellenONE[28,29], and it supports gentler and more robust isolation than FACS, which is particularly helpful with primary cells[18].

When analyzing the proteomes of single cells from tissues, the spatial context should be characterized as best as possible, including both the location of each cell in the tissue and the extracellular matrix around it. Although a great area of interest, such single-cell MS proteomics analyses are in their infancy. Feasible approaches for spatial analysis include tissue sectioning by cryotome and laser capture microdissection (LCM), which can be used to extract individual cells[30]. LCM has been used for spatially resolved extraction and subsequent MS analysis of tissue regions[21,22]. While such small samples can be physically isolated, MS analysis of individual cells or organelles extracted by LCM is challenging and has not yet been achieved. We recommend avoiding the use of protocols that require clean-up from detergents for tissue disruption, and instead preferring methods using only MS compatible reagents.

**Reducing contamination.** Minimizing sources of contaminating ion species that disproportionately affect the analysis of small samples is critical for single-cell proteomic measurements. Contaminating ions can result from many sources, including reagents used during sample preparation, impure solvents, extractables and leachables from sample contact surfaces, and especially carry-over peptides from previous single-cell or bulk runs that may persist within liquid-handling instrumental components, capillaries, and stationary phases, such as needle washing solutions and column-retained analytes in liquid chromatography (LC) and reservoirs in capillary electrophoresis (CE). Typically, only about 1% of peptides persist on C18 column resin following a run, and they may appear in subsequent runs as carry-over "ghost" signal[31]. Fortunately, these carry-over peptides generally make a quantitatively insignificant contribution to consecutive samples of comparable amounts. However, when bulk samples are interspersed with single-cell runs, carry-over peptides from these bulk samples may significantly contaminate or even dwarf the peptide content derived from the single cells. Thus, contaminants from bulk sample runs are often incompatible with quantitative single-cell analysis on the same LC-MS system. Before analyzing single-cell samples, analytical columns must be evaluated rigorously and deemed free of carry-over, as previously described[5,32]. Other non-peptidic contaminants, such as leached plasticizers, phthalates, and ions derived from airborne contaminants, often appear as singly charged ions and can be specifically suppressed by ion mobility approaches[7,32–34] or, in the case of airborne contaminants, by simple air-filtration devices (e.g., ABIRD)[5,35].

Because the ratio of sample preparation volume to protein content is significantly increased, the amount of reagents to protein content is also significantly increased when preparing single cells individually. Thus, reducing sample preparation volumes mitigates the effect of contaminant ions originating from reagents such as trypsin ions of mass tags[2,36]. Indeed, reducing sample preparation volumes to 2-20nl proportionally reduces reagent amounts per single cell compared to multiwell based methods, which in turn reduces the ion current from singly charged contaminant ions[6].

**Sample preparation.** Ideally, sample preparation should consist of minimal steps designed to minimize sample handling, associated losses, and the introduction of contaminants. For bottom-up proteomic analyses, workflows must include steps of cell lysis/protein extraction and proteolytic digestion. Given the picogram-levels of protein present in a single cell, it is crucial to minimize contaminants and maximize sample recovery for downstream analysis. Fortunately, the composition and geometries of single cells isolated from patients and animals lend themselves to disruption under relatively gentle conditions, such as a freeze-heat cycle[5,37,38] or nonionic surfactants[39,40]. Such clean lysis methods are preferable over MS-incompatible chemical treatments (e.g., sodium dodecyl sulfate or urea) that require loss-prone cleanup before MS analysis[41]. It can be beneficial to miniaturize processing volumes to the nanoliter scale to minimize exposure to potentially adsorptive surfaces[2,6], although such approaches may have limited accessibility. In contrast, sample preparations using low microliter volumes offer broadly accessible options[16,37,42] and are described in detailed protocols[5,38]. Regardless of the selected preparation workflow, it is recommended that cells be prepared in batches that are as large as possible to minimize technical variability in sample handling. To this end, several liquid handling tools have been successfully coupled with single-cell proteomics workflows to increase throughput and reduce technical variability. In particular, the Formulatrix mantis and the open-trons have been adapted for 384-well plate based sample preparation[5,37,42]. The CellenONE system has also been employed for several automated protocols using microfabricated multi-well chips[2,28,43], or using droplets on glass slides[29]. We expect this landscape to continuously evolve towards increased consistency and throughput of sample handling.

**Maximizing sample delivery to mass analyzers** For sample-limited analyses, it is especially important to maximize ionization efficiency (the fraction of gas-phase ions created from solution-phase molecules) and the transmission of those ions to the mass analyzer. Lower volumetric flow rates produce smaller, more readily desolvated charged droplets at the electrospray source, leading to increased ionization efficiency[44,45]. As such, reducing the flow rate of separations from hundreds to tens of nanoliters per minute can increase measurement sensitivity, but currently these gains must be achieved with custom-packed narrow-bore columns and may compromise robustness and measurement throughput[20]. Maximizing separation efficiency is also important, as narrower peaks increase the concentration of eluting peptides and simplify the mixture entering the mass spectrometer at a given time[46]. A number of commercial nanoLC systems and columns provide a reasonable combination of sensitivity and efficiency for single-cell proteomics and these are recommended for most practitioners. Alternative high resolution separation techniques employing orthogonal separation mechanisms, e.g., capillary electrophoresis and ion mobility as well as multidimensional techniques may potentially be employed as front-end approaches in MS-based single-cell proteomics[11,47,48]. Increasing ion transmission in the mass spectrometer is generally the purview of instrument developers and companies, and future gains in this area are expected to further benefit single-cell proteomics. Lastly, when injecting samples for analysis by LC-MS, because of the low protein amount, it is often desirable to inject the entire sample. If the samples are resuspended in too small a volume, the autosampler may miss portions of the sample or may inject air into the lines, which adversely affects chromatography. Thus, we recommended striking the correct balance of suspension volume that prevents air injections and maximizes sample delivery. This balance depends partially on the autosamplers, sample vials, their shape and size. One implementation shown to perform robustly includes injecting 1 microliter samples from 384-well plates[5,6,18].

**Controls.** Experimental designs should provide an estimate of quantitative accuracy, precision and background contamination. Precise measurements may arise from reproducing systematic biases, such as integration of the same background contaminants. Measurement precision can therefore be assessed by repeat measurements. In contrast, benchmarking measurement accuracy requires positive controls, i.e., proteins with known abundances. One approach to benchmarking is incorporating into the experimental design samples with known quantitative values to assess quantitative accuracy. These controls may be derived from independent measurements based on fluorescent proteins or well-validated affinity reagents. Other positive controls include spike-in peptides[18], proteins, or even proteomes in predefined ratios as performed for LFQbench experiments[49]. When cells from clusters consisting of different cell types can be isolated, the relative protein levels of the isolated cells may be quantified with validated bulk assays and used to benchmark the in-silico averaged single-cell estimates, an approach used by multiple studies[5,9,16,18,29]. A positive control for sample preparation may include bulk cell lysates diluted to the single-cell level. Estimating protein amounts corresponding to single cells is challenging, and thus we recommend starting with cell lysate from precisely known cell numbers (e.g., estimated by counting cells with a hemocytometer) and performing serial dilution to single-cell level[5]. Negative control samples, which do not contain single cells, should be processed identically to the single-cell samples. Such negative controls are useful for estimating cross-labeling, background noise and carry-over contaminants.

When matching between runs (MBR) is used to propagate sequence identification, MBR controls should be included. Empty samples contain few ions, if any, that may be associated with incorrect sequences. Thus, using empty samples may lead to underestimating MBR false discoveries. MBR may be evaluated more rigorously by applying it to samples containing either mixed species proteomes or single species proteomes and then estimating the number of incorrectly propagated species. Such MBR controls (samples of mixed yeast and bacteria proteomes or only yeast proteomes) have been used to benchmark sequence propagation within a run[7], and similar standards should be used for benchmarking MBR. While MBR is best evaluated in each study with samples designed to reflect the analyzed proteomes, the field may benefit from preparing a community reference samples that were analyzed in multiple laboratories and used for benchmarking MBR algorithms.

**Batch effects.** Systematic differences between groups of samples (biological) and analyses (technical) may lead to data biases, which may be mistaken for cell heterogeneity, and thus complicate results interpretation or sacrifice scientific rigor. To estimate and correct batch effects, treatments and analytical batches must be randomized whenever possible[50]. We recommend that treatment and batches are randomized, so that batch effects can be corrected (estimate and remove batch effects from data) or modeled (e.g., include batch effect as a covariate in models). When randomization is not performed, biological and technical factors may be fundamentally inseparable. For experiments wherein randomization was not performed, downstream statistical analyses should include the batch information as covariates. These considerations are similar to those for bulk experiments, which have been previously described[51]. Furthermore, we recommend that all batches include the same reference sample, which can be derived from a bulk sample diluted close to a single-cell level.

**Statistical power.** Studies should be designed with sufficient statistical power, which depends on the effect sizes, on the measurement accuracy and precision, and on the number of single cells analyzed per condition. Simple experiments with large effect sizes, such as analyzing different cell lines, can achieve adequate statistical power with a few dozen single cells. Such experiments were common as proof of principle studies demonstrating analytical workflows. In contrast, experimental designs including primary cells, smaller effect sizes (e.g., protein variability within a cell type[6]), multiple treatment groups or patient cohorts, require a much larger number of single cells and patients to achieve adequate statistical power[52–55]. Thus, increasing the throughput of single-cell proteomics and incorporating power estimates in the experimental designs will become crucial as single-cell proteomics matures[56,57].

**Methods for MS data acquisition.** Existing methods can be grouped into label-free, which analyze one cell per sample, and multiplexed, which analyze multiple cells per sample. Label free methods benefit from simpler sample preparation, while multiplexed methods benefit from analyzing more cells per unit time[5]. When multiplexing is performed by isobaric mass tags, quantification is adversely affected by the co-isolation and co-fragmentation of precursors. This co-isolation can be mitigated by targeting the apexes of elution peaks and using narrow isolation windows[16,18]. The co-isolation artifacts on quantification can be overcome by performing quantification on peptide-specific and sample-specific ions, as in the case of plexDIA, which multiplexes cells with non-isobaric mass tags[7,58]. Isobaric mass tags have been used in combination with a carrier sample, which reduces sample losses and facilitates peptide sequence identification[59]. This approach has raised concerns as high carrier amounts may allow confident peptide identification without sampling sufficient peptide copies from the single cells to achieve precise quantification[60,61]. To address these concerns, multiple groups have converged on guidelines for balancing the precision and throughput of single-cell analysis using isobaric carriers[60,61]. Specifically, we recommend using carrier amounts and instrument settings (target AGC) that allow accumulating ions for MS2 scans for the intended time (e.g., at least 100-200ms) and verifying method performance by inspecting the distributions of accumulation times and S/N ratios for reporter ions.

We recommend, when possible, cross-validating protein measurements with different methods. Often, such cross-validation may be performed using the same MS instruments, and the results directly reported and compared in the same paper. Such cross-validation studies are particularly useful for supporting new and surprising biological results. As an example, Leduc et al.[6] observed a gradient of phenotypic states and protein covariation within a cluster of melanoma calls not primized for drug resistance. The authors cross-validated these observations by analyzing biological replicates of the melanoma cells both by isobaric multiplexing with pSCoPE[18] and by non-isobaric multiplexing plexDIA[7]. The results from the two methods were directly compared and reported in parallel so that the degree of biological and technical reproducibility can be evaluated[6]. Cross-validation analysis can also benefit from using different sample preparation methods or enzymes for protein digestion. In such cross-validation analyses, quantitative trends supported by multiple methods and biological replicates are more likely to reflect biological signals rather than method-specific artifacts.

**Method selection and optimization.** The MS methods and their parameters should be selected depending on the priorities of the analysis. Maximizing the number of cells analyzed is best achieved with short separation times and multiplexed methods[57]. Maximizing the proteome depth is best achieved with longer separation methods, while maximizing the number of copies sampled per protein is best achieved with MS1-based methods and longer ion accumulation times[7,36]. Multiple objectives, such as increased consistency, dynamic range and coverage, may best be simultaneously optimized with intelligent data acquisition strategies[18,36,57,62]. The size of the isobaric carrier used can also help emphasize project priorities, such as depth of proteome coverage versus copy numbers sampled per peptide[60,61]. Choosing optimal method parameters can be time consuming, and software for systematic, data driven optimization can speed up such optimizations[63].

# Data evaluation and interpretation

**Defining and evaluating reproducibility.** We begin discussing data reproducibility and evaluation by briefly defining several levels of increasing difficulty, namely *repeating*, *reproducing*, and *replicating*[64]. Repeating a computational experiment or an analysis simply consists in using the exact same data, code, software and environment (typically the same computer), assuming that these are still available. Reproducing an experiment or analysis is an attempt by a different person that will mimic the original setup by downloading data and code, without necessarily having access to the same software environment. Replication represents a further challenge where the results are to be obtained using new code/implementation/software; it is only possible with extensive and detailed description of the performed analyses. This description must include the versions of all software and databases used as well as all search parameters, ideally saved as structured documents, e.g., xml.

The advent of containerised workflows now facilitates analysis replication without the need to go through the complicated process of setting up the exact same computational pipeline manually[65]. Still, many data analysis solutions, such as the R computational environment (R core team, 2021), are typically used in a non-containerised fashion, making the end-results potentially version-dependent and difficult to reproduce, especially in the long term. Thus, we recommend introducing workflow managers to facilitate the reproduction of single-cell proteomics data analysis. Of note, while the different reproducibility concepts are often described in the context of computational experiments and data analysis, they can also be extrapolated to experimental workflows. For example, detailed, dynamic, and version-controlled protocols, such as those on the protocols.io platform, can facilitate experimental reproduction. Other containered options, such as Docker images, are a fruitful strategy for standardizing computational analysis of complex datasets, with cross-platform compatibility and robust version control.

**Batch effects and cellular uniqueness**. Two factors should be considered when reproducing single-cell protein measurements. First, no two cells are identical. Thus, we may reasonably hope to reproduce clusters of cells and trends (such as protein abundance differences between cell types or cell states), but not the exact molecular levels for each analyzed cell. Second, batch effects may increase the apparent level of reproducibility (when biases are shared between replicates, such as peptide adhesion-losses or co-isolation) or decrease it (when biases differ between replicates, such as protein digestion biases). Thus, assessments and reports of reproducibility need to be specific about precisely what is being reproduced and how this may be impacted by batch effects originating from all steps, from cell isolation to data processing.

**Evaluating quantitative accuracy.** Quantitative accuracy is a measure of how closely the measurements correspond to known true values, as in the case of proteomes mixed in experimenter-determined ratios, Fig. 2a. When the true abundances are not known, evaluating accuracy is not possible and is sometimes confused with repeatability or precision. Yet, these quantities can be quite different as illustrated in Fig. 2a. Similarly, high correlation between replicates may be interpreted as evidence that the measurements are quantitatively accurate. This interpretation is wrong: many systematic errors may lead to erroneous measurements that are nonetheless very reproducible. Thus, reproducibility alone is insufficient to evaluate data quality. Because single cell proteomics pushes the limits of sensitivity for MS-based measurements, the quality of measurements depends on the number of ions measured from each single-cell population[60,61]. For example, if too few ions are sampled, the stochasticity of sampling results in counting noise, i.e., low precision estimates and technical variation in estimated protein abundances, which should be clearly distinguished from biological variability[36]. Such counting noise also affects single-cell RNA-seq methods (that sample even fewer RNA copies per cell), and some of the models developed for RNA-seq data may help handle counting noise in MS data as well[66]. Mixing ratios of 1:1 can be used to evaluate ion sampling and precision but not accuracy since this ratio is not sensitive to systematic biases, such as co-isolation and interference. Accuracy can be evaluated relative to ground truth ratios, as created by mixing the proteomes of different species in known ratios[7,49,67]. As another approach, measurements of relative protein abundance by established bulk methods can provide useful benchmarks for evaluating corresponding single-cell-level measurements[7,9,16]. On a smaller scale, accuracy may be estimated for a limited number of proteins by spiking corresponding peptides at known ratios or using measurements that are as independent as possible; such independent measurements include fluorescent proteins whose abundance is measured fluorometrically[1] or immunoassays with high specificity, such as proximity ligation assays that enhance specificity by using multiple affinity reagents per protein[68].

Quantitative precision and accuracy are different, and their importance is highly dependent on the analysis. For example, cell clustering benefits from high precision measurements and may tolerate low quantitative accuracy. In contrast, protein covariation analysis[6,19] and biophysical modeling[12] are more dependent on quantitative accuracy. Thus, benchmarks should clearly distinguish between accuracy and precision and focus on the metric that is more relevant to the biological goals of the analysis.

Comparisons between absolute protein intensities conflate the variance due to protein abundance variation across the compared samples (conditions) and across different proteins and may result in misleading impressions[69]. For example, the high correlation between the proteomes of T cells and monocytes in Fig. 2b may be interpreted as indicating that the two proteomes are very similar. Yet, many proteins differ in abundance reproducibly between T cells and monocytes, Fig. 2c. Thus, correlations between estimates of absolute protein abundance should not be used as benchmarks for relative protein quantification.

**Evaluating quantitative consistency.** Outside of carefully designed benchmarking experiments, the true protein abundances are unknown, and thus the accuracy of quantification cannot be directly benchmarked. However, it is often possible to evaluate the reliability of MS measurements based on comparing the quantitative agreement between (i) different peptide fragments from the same peptide (Fig. 2d) or (ii) different peptides originating from the same protein. For example, the internal consistency of relative quantification for a peptide may be assessed by comparing the relative quantification based on its precursors and fragments, as shown for single-cell plexDIA data in Fig. 2d. The degree of (dis)agreement may be quantified by the coefficient of variation (CV) for these estimates. Similarly, the CV estimated from the relative levels of different peptides originating from the same protein may provide a useful measure of reliability. This analysis is limited by the existence of proteoforms[70,71] but nonetheless may provide useful estimates of data quality. Note that this CV is very different from the CV computed using absolute peptide intensities or the CV computed between replicates. In the latter case, when comparing CVs across different analytical or experimental conditions, it is imperative to account for varying dataset sizes; i.e. a rigorous comparison between experimental methods would rely on peptides and proteins identified identified and quantified across all samples, rather than also including peptides and proteins identified uniquely in individual experiments[63].

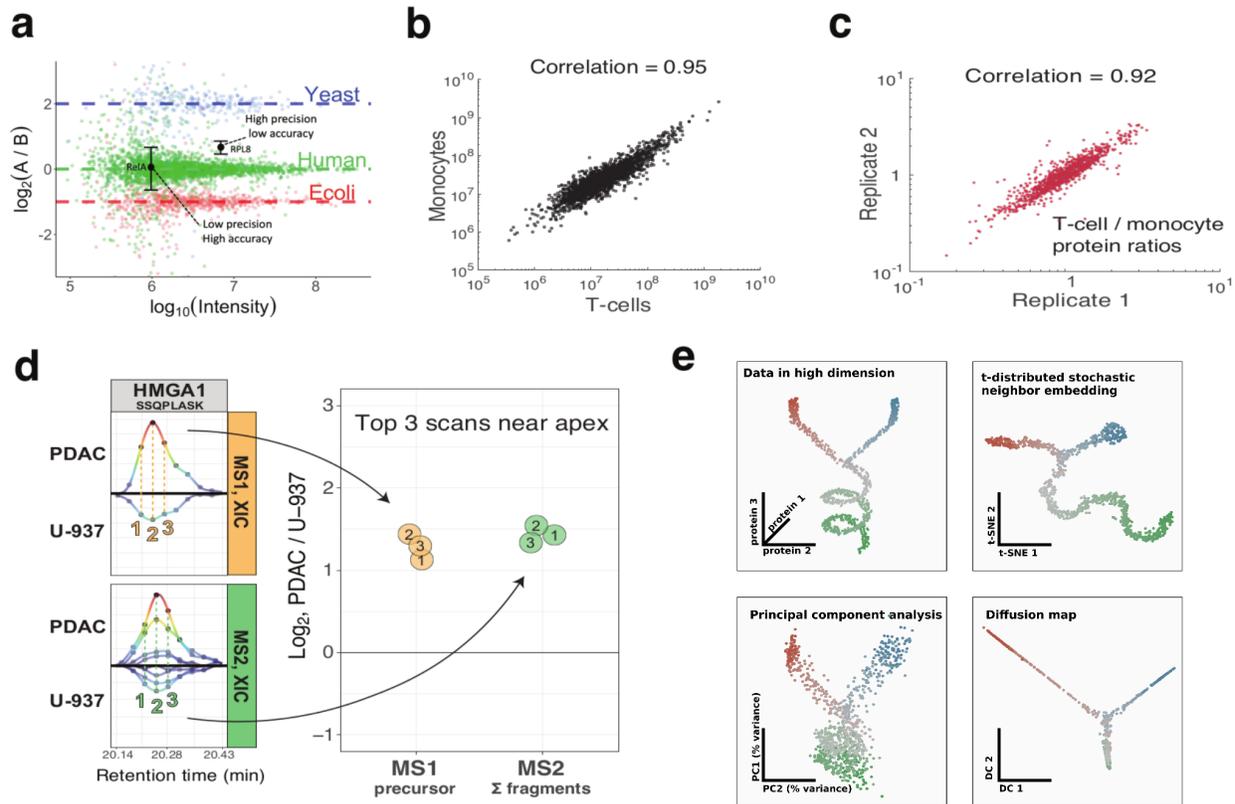

**Figure 2 | Evaluating and interpreting single-cell proteomics data. a**, Quantitative accuracy of protein ratios between samples A and B measured by label-free DIA analysis relative to the corresponding mixing ratios denoted by dotted lines[7]. Some proteins are quantified with high precision but low accuracy (e.g., RPL9), while others are quantified with high accuracy and low precision (e.g., RelA). The proteome of T-cells and monocytes correlate strongly **(b)** despite the fact that many proteins are differentially abundant between the two cell types **(c)**. Data for panels b and c are from Specht et al.[37] **d**, Extracted ion current from single-cell MS measurements by plexDIA. Such data allow quantifying peptides at both MS1 and MS2 levels, which can be used to evaluate the consistency and reliability of the quantification. This example data from Derks et al.[7] show that relative levels estimated from precursors (peach color) agree with the relative levels estimated from the corresponding summed up fragments (green color). At both MS1 and MS2 levels, 3 estimates are obtained based on the 3 scans closest to the elution peak apex. **e**, Different dimensionality reduction methods approximate the data in different ways. We simulated 3D data for 3 cell states, where one cell state (green) progressively diverges to 2 distinct cell states (blue and red, top left panel). Projecting the data to 2D loses information. Specifically, PCA loses the non-linear cycling effect and mixes early (green) and intermediate (gray) cells, tSNE does not correctly capture the distances between the 3 populations and diffusion maps do not capture the noise in the data and compress the early state cells. The code for this simulation is available at https://github.com/SlavovLab/SCP_recommendations.

**Accounting for biological and technical covariates.** Single cells differ in size and thus protein content. Consequently, cell size is a major confounder for the differences in protein intensities between cells[6]. The basic normalization strategy here consists of subtracting from log-transformed protein quantities the respective medians across the proteins quantified[16]. However, this normalization can be undermined if the subset of quantified proteins varies significantly across single cells. Such variation may stem from differences in total protein amounts between cells or experimental variability, which may lead to differences in the numbers of missing values and proteins accurately quantified. In case of such variation, normalization should be based on a common subset of proteins, or against a common reference, as described by Franks *et al.*[69]. Thus, the processing of single-cell MS proteomic data is likely to be improved in the future with the development of more advanced normalization strategies, which may build upon those developed for scRNA-seq experiments[72] to mitigate similar challenges. To compensate for imperfect normalization, we suggest including a variable representative of the cell size, such as total protein content estimated from LC-MS data or forward scatter from FACS, as a covariate in downstream analyses.

**Managing missing data.** One of the common challenges in analyzing single-cell data is handling the presence of missing values[50,73]. These tend to be more prevalent in single-cell proteomics, compared to typical bulk experiments since some proteins may be below the limit of detection (especially in smaller cells) or may not be sent for MS2 analysis in every single cell Indeed, single-cell proteomics operates at the boundary of sensitivity of LC-MS instrumentation, and therefore a peptide quantified in some cells is below the detection limit in other cells. The missing data is a source of uncertainty that should be propagated through the analysis and ultimately reflected in the final conclusions. Many analyses may be conducted using only the observed data (without using imputed values), which assumes that the observed data are representative of the missing data. Yet, a common strategy for dealing with missing data is to impute missing values prior to any downstream analysis. Ideally, imputation should take into account the nature of missing data (e.g., missing at random or not at random[74]) in determining appropriate imputation methods. The type of missigness is determined by the mechanism leading to missing values, which depends on the algorithm for peptide sampling during mass spectrometric analysis. Shotgun methods using the topN heuristic introduce missing values that are more likely to occur at random, as they originate from the stochastic selection of precursors for MS2 scans. In contrast, DIA and prioritized methods send precursors for MS2 scans deterministically and most missing values likely correspond to peptides below the limit of detection rather than missing at random. Comprehensive imputation methods for single-cell proteomics are yet to be developed and benchmarked, but recommendations developed for bulk proteomics methods may serve as useful guides[74–76]. While some recently developed methods for scRNA data may be adapted to proteomics, ultimately, the field needs methods which are specifically tailored to the mechanisms leading to missing peptides and proteins. Multiple imputation can be used to quantify the uncertainty in the results for a given missing data method. Although computationally demanding, it is also prudent to impute using different missing data models to further characterize the sensitivity of the results to unverifiable assumptions about the missingness mechanism. A simple example of this strategy would be to perform downstream data analysis, such as PCA, on the imputed data and compare the results to the analysis performed on the unimputed data[16,18]. Results that are insensitive to different types of imputation models are more reliable, while those that are contingent on the validity of a particular assumption about missingness should be viewed with more skepticism.

**Dimensionality reduction.** High-dimensional single-cell data are often projected onto low-dimensional manifolds to aid visualization and to denoise the data. While such projections can be useful, the reduced data representations are incomplete approximations of the full data and often lose aspects of the data, as illustrated in Fig. 2e by projecting a 3D dataset into different 2D projections. As such, different low-dimensional projections may selectively highlight certain aspects of the data while obscuring others, Fig. 2e. At worst, they may severely distort the original data[77]. Thus, we recommend using dimensionality reduction as an initial data analysis step that requires further scrutiny. Conclusions derived from reduced data representations, such as clustering of cells, should be validated against the high-dimensional data. The validation can be as simple as computing and comparing the distances between the cells in a higher dimensional space, as demonstrated with macrophage clusters defined based on single-cell RNA and protein data[78].

While dimensionality reduction representations can be useful for visualization, clustering of cell types in low-dimensional manifolds is inadequate for benchmarking quantification. Such representations indicate whether the cells cluster in a low-dimensional space, but they indicate little about the factors, whether biological or technical, that could be driving the clustering. More fundamentally, low-dimensional data reductions often account for only a fraction of the total variance in the data and thus may exclude relevant sources of biological variability, Fig. 2e. Some methods, such as PCA, better preserve global distances and are thus more amenable to interpretation, as opposed to their non-linear counterparts, such as t-distributed stochastic neighbor embedding (tSNE)[79] or uniform manifold approximation and projection (UMAP)[80]; in these two latter methods, the separation between cell types is sensitive to various tuning parameters, which may introduce subjectivity. Furthermore, only the small distances within clusters are interpretable. Thus, when results, such as cluster assignment, are based on a low-dimensional manifold, we additionally recommend showing the corresponding distances in higher dimensional space, e.g., as distributions of pairwise distances between single cells within and across clusters[78].

When dimensionality reduction is used for clustering cells, we recommend including positive controls. These controls may be bulk samples composed of purified cell types (if such isolation is possible) from the same population as the single cells of interest. Such positive controls should be prepared in tandem with the single cells. Then, both positive controls and single cells can be projected simultaneously on the low-dimensional manifold. This type of analysis provides useful evidence for evaluating the clustering[16,18] patterns: The degree to which the positive controls and the single cells of the same type cluster together indicates the consistency of the measurements. To further determine if sample preparation is driving any clustering, we also recommend evaluating whether principal components correlate with technical covariates (such as batches, missing value rate, or mass tags) and correcting for these dependencies if needed.

**Managing and propagating uncertainty.** As discussed above, assumptions about missing data and the application of dimensionality reduction methods can substantially influence the final conclusions. Thresholds, such as filters for excluding single cells due to failed sample preparation or for excluding peptides due to high levels of interference can also influence the results[16,50]. Such choices should be based on objective grounds, such as true and false discovery rates derived from controls. For example, negative controls allow establishing objective filters for failed single cells as already implemented in multiple pipelines[7,16,50]. When thresholds are set based on subjective choices, this should be explicitly stated, and the choices treated as a source of uncertainty in the final results. When possible, the sensitivity of the results to all experimental and methodological choices should clearly be conveyed.

*Cross-validation of single-cell proteomics measurements*

Single-cell proteomics results may be evaluated and bolstered by using different methods for sample preparation and data acquisition, ideally methods that share minimal biases. Each method has its own set of biases, which are difficult to overcome and may induce systemic artifacts that contribute to the final interpretation of the data. To safeguard against such situations, different methods that have divergent technical biases should be used to obtain single-cell data. For example, plexDIA and pSCoPE use two very different types of tags (non-isobaric vs isobaric) and perform quantification at different levels and on different ions (MS1 quantification of precursors vs. MS quantification of reporter ions); thus, using these methods to analyze single-cell proteomes can cross-validate the final biological results[6]. In such cases, the protein quantification results supported by multiple methods with different quantification biases are more likely to reflect biological, rather than technical, factors.

*Interpreting features of single-cell proteomics data*
Algorithms underlying peptide identification have evolved along with technological advances in data generation to use the increasing set of features from bulk proteomic data. Features measured at the single-cell level may differ significantly from corresponding bulk samples since lowly abundant fragments may not be detected and other fragments may have lower signal relative to background noise[81]. Mitigating these challenges may benefit from directed efforts dedicated to developing robust models trained on features that have the greatest discriminatory power at single-cell level input. These models may incorporate additional features with search engine results, as implemented by Mokapot[82] and DART-ID[83]. To guard against false positive identifications, we recommend scrutinizing any peptides identified in single-cell samples but not identified in larger bulk samples from the same biological systems. Such identifications are unlikely to be correct, especially for DIA experiments, and the spectra supporting them (e.g., extracted ion current) should be examined and data analysis methods reassessed.

To improve proteome coverage, new search engines may be designed and optimized to exploit regular patterns in the data, such as the precisely known and measured mass shifts in the precursors and fragments of plexDIA data[84,85]. Indeed, current single-cell proteomic mass spectrometry methods are capable of measuring tens of thousands of peptide-like features; however, only a small fraction (between 1-10%) of these features are assigned sequences at 1% FDR[20,61,84]. Anticipated models that successfully address these unique challenges will enable identification rates to approach those of bulk experiments and extend the utility of single-cell proteomics in biomedical research[84].

# Reporting standards

The goal of reporting is to enable other researchers to repeat, reproduce, assess, and build upon published data and their interpretation[86]. While reproduction and replication do not guarantee accuracy, they build trust in the analysis process through verifiability, thus strengthening confidence in the reported data and results. Replication requires sufficient documentation of the metadata, and a good starting place for reporting metadata are formats developed for bulk MS data[22,87], including specifically for proteomics data[88], those prepared by journals[89–92] and societies[93], as well as for single-cell RNA-seq data[94]. Nonetheless, single-cell MS proteomics data have additional aspects that need to be reported, which are the focus of our recommendations. Below, we document the essential information needed to provide value to single-cell proteomics data, meta-data, and analysis results.

**Experimental design**. The detailed design of the experiments should be reported, which includes treatment groups, number of single cells per group, sampling methods, and analysis batches, Fig. 3. The experimental design should be reported as a table listing each analyzed single cell on its corresponding row and each descriptor in its corresponding column. Specifically, columns document biological and technical descriptors, i.e. variables that describe the biology of the measured cells and technical factors that are likely to influence the measurements. Required biological descriptors contain sample type (such as single cell, carrier, empty, or control sample) and biological group, such as treatment condition or patient/donor identifier, cell line, organism and organ/part of origin (if cells from multiple organisms or multiple organs are assayed) and biological characteristics for multi-sample and/or multi-condition studies. When available, additional biological descriptors may include the cell type and/or cell state (e.g., their spatial and temporal information in tissues), physical markers (e.g., pigmentation, measured by FACS), cell size, and aspect ratio. These descriptors apply only to single-cell samples and thus will remain empty for some samples, such as negative controls. Note that some of these descriptors might be known before the acquisition of the data (such as cell types based on different cell cultures or following from FACS sorting) or be the results of downstream analyses (such as cell types or cell states inferred from clustering or differential abundance analysis). Required technical descriptors include the raw data filenames and acquisition dates, as well as variables that describe the underlying technical variability, whether it is expected to be significant compared to the biological variability or not. These descriptors include all batch factors related to cell isolation, sample preparation, peptide and protein separation (chromatography or electrophoresis batches), operator(s) and instruments (when multiple mass spectrometers were used), as well as chemical mass tags/labels (in case of labeled quantitation, e.g., TMTpro). Such a sample metadata table is also a simple and valuable quality control tool since it allows for verification that the number of rows in the table matches the number of cells reported in the paper and that the number and names of raw data files extracted from the table is compatible with the files in the data repositories (see below). We encourage researchers to document additional descriptors where needed, such as variables defining subsets of cells pertaining to distinct analyses. This sample metadata table should be complemented by a text file (often called README) that further describes each of these descriptors and the overall experiment. We include a standard README file and recommend using it to facilitate standardization and data reuse. The README file should contain a summary of the study design and the protocols. The measurement units of some of the descriptors (such as micrometers for cell sizes) should also be documented in the README file, as opposed to encoding them as a suffix in the descriptor's name.

Ideally, the raw and processed MS data should be shared using open formats, such as HUPO Proteomics Standards Initiative community-developed formats dedicated to mass spectrometry data: mzML[95] for raw data, mzIdentML[96] for search results, and mzTab[97] or text-based spreadsheet for quantitative data. When binary formats from proprietary software are provided, they should be converted into an open and accessible format as well when possible. Raw data files and search results should be made available through dedicated repositories, such as PRIDE[88] and MassIVE[98]. Code repositories, such as GitLab or GitHub[99], are ideal to store and share code, scripts, notebooks and, when size permits, quantitative data matrices. When these become too large to be stored directly with the scripts that generate them, they should be made available in institutional or general-purpose open repositories, such as Zenodo or Open Science Framework, or on publicly available cloud storage. The latter however requires a commitment by the data provider to keep the data public. The README file (an example is provided as Supporting File 1) containing the description of the experimental design and the different locations holding data should be provided in all these locations. The manuscript material and method section and/or the supplementary information should provide the experiment identifiers and links to all the external data and metadata resources. Editors and reviewers should systematically require the deposition of all data, metadata, and analysis details as a condition for paper acceptance and publication.

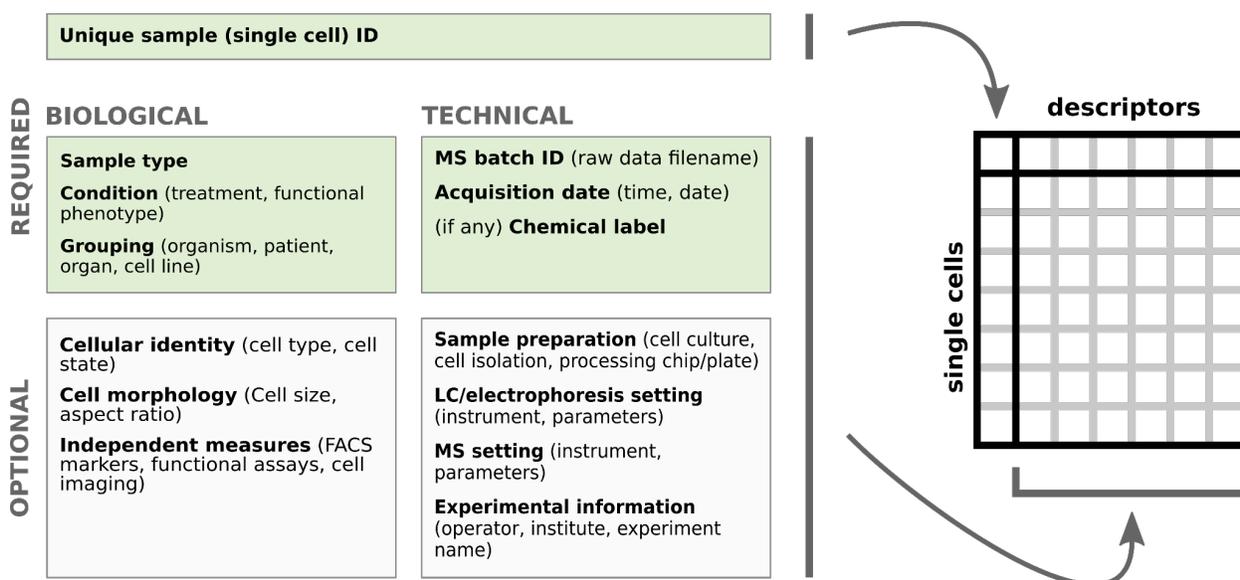

**Figure 3 | Suggested descriptors of single-cell proteomics samples.** The metadata should include the experimental design table with rows corresponding to single cells and columns to the required and optional features listed here (an example is provided as Supporting File 2). Attributes provided in parenthesis are given as examples or for clarification. The green frames highlight required descriptors, while the gray ones include a non-exhaustive list of optional descriptors, which may also include spatial (e.g., position in tissues) and temporal information for the cells when available. The descriptors (and their units, when relevant) should be documented in the experiment's dedicated README file.

While these data sharing requirements apply broadly to proteomics experiments, some are specific to single-cell proteomics (such as single-cell isolation), and some are made more important because of the aim to analyze tens of thousands of single cells per experiment[57]. Such sample sizes are required to adequately power the analysis of dozens of cellular clusters and states across many treatment conditions and individuals. The large sample sizes, in turn, considerably increase the importance of reporting batches, including all variations in the course of sample preparation and data acquisition, as well as the known phenotypic descriptors for each single cell. These reporting requirements expand the essential descriptors in the metadata. Large study sizes also heighten the importance of reporting datasets from intermediate processing steps, such as search results and peptides x cells matrices, to reduce the computational burden on reproducing individual steps from the analysis.

BOX: Despite its apparent simplicity, file naming deserves thoughtful consideration. Files names should be unique (unlikely to be used in other studies) and linked to the measurements in the file; additional good practices are summarized in ref.[100]. We suggest thinking about file naming and file naming conventions to easily identify groups of files pertaining to specific meta-data elements or experiments. A systematic file naming convention allows files to be both machine and human readable and searchable. File names should avoid using any special characters and use the same character (such as a dash or an underscore, rather than spaces) to separate the different elements of the file names. If using dates to list files chronologically, the YYYYMMDD format should be used. Finally, these naming conventions and any abbreviations used as part of the filenames need to be documented in the main README file.

Sharing the data is necessary but insufficient for replication data reuse. Any analysis of the data is likely to require the associated metadata. Furthermore, the exact processing of data has to be documented and shared as it can profoundly influence the final results that are used to infer biological interpretations. Data processing can hardly (and should not need to) be retro-engineered from the result files. Therefore, annotated scripts or notebooks used to process, prepare and analyze the data need to be provided with the data. Using software for single-cell proteomics, such as the *scp* R/Bioconductor package[50,101], the *sceptre* python package[9], the SCoPE2 pipeline[16,102], or the Scripts and Pipelines for Proteomics (SPP)[103] can help standardize workflows across laboratories. Packages that allow comparing structured and repeatable data processing, including evaluating different algorithms for a processing step, provide further advantages[50,101]. Software platforms that support exporting the commands and parameters used should be strongly preferred since audit log and/or parameter files can help tracking and later reproducing the different processing steps, including software and the versions used at each step. We strongly advise against using non-reproducible software given the difficulty in capturing their operation. Given the rapid evolution of the field, specific description of the methods should be favored over simply referring to other publications using 'as previously analyzed in [ref]'. When reporting results, it should be made clear which data the result refers to. This is, for example, crucial when reporting coefficients of variation (CVs), where CVs on log-transformed data are lower than on the linear scale. CVs can be used to quantify very different quantities, such as repeatability between MS runs or consistency of protein quantification based on different peptides, and thus the exact quantity must be explicitly specified. Similarly, researchers should systematically report major features of the data that influence the results and how these were observed and addressed throughout the data analysis. These typically include missing values and batch effects. Reproducibility requires to go beyond the minimalist 'Material and method' sections that often fail to describe the processing of samples and data to enable their replication.

Often, studies include several sets of raw, identification, and quantitation files, addressing different research questions, such as different instruments or MS settings, different cell types or growth conditions, and different subjects. A single dump or all files makes data reuse challenging. In such situations, it is advisable to split the file in different folders, following a consistent structure. The high-level README file, already mentioned above, should describe what each of these folders correspond to, and each folder should contain its own README file describing its content in detail and the specific points these sets of files aim to address.

As described above, data acquisition strategies are inextricably linked to both the number of proteins quantified and the quality of quantitation in single-cell proteomics experiments. While the reporting of MS acquisition details is not necessarily required for reanalysis of the data, acquiring similar data could be impractical or impossible if key details are not reported. This is even more evident with the rise of intelligent data acquisition strategies that often have more advanced, non-standard parameters or use third party (non-vendor) supplied software. Luckily, most raw data files report the parameters used for analysis and some vendors have enabled method generation from a raw data file. However, for instances where third-party software makes real-time decisions that alter mass spectrometer operation - the software should be made available to the broader research community. Ideally this software would be open source. If it needs to be delivered as a compiled executable, the underlying algorithms should be described in such a way that others could reproduce a similar method. Furthermore, the reporting of parameters relevant to the decisions made in real-time as well as the output of real-time decisions would ideally be provided. These considerations would enable faster implementation in laboratories trying to replicate published results on their own instrumentation.

These reporting guidelines might give the impression that a lot of additional work is expected when reporting on studies according to our recommendations, many of which apply to all proteomics studies. Yet, the recommendations merely highlight good scientific practice, to be implemented continuously, starting when the research is designed, when the data are acquired, processed, and eventually interpreted. When so implemented, they become habits enabling robust research rather than a burden to be addressed at the end of the research project. Data, meta-data, and analyses documentation and reporting happen at different stages of the analysis process and rely on each other. The investment that we are suggesting here is simply work that is spread across the research project, rather than extra work done at the very end of it[104].

# Conclusions and perspectives

The adoption of these guidelines by the scientific community and their promotion by journals and data archives is essential for establishing solid foundations for the emerging field of single-cell MS proteomics and to uphold scientific rigor. The suggested reporting standards will facilitate all levels of replication and thus promote the dissemination, improvement, and adoption of single-cell technologies and data analysis. Sound data evaluation and interpretation will further promote the reuse of single-cell proteomics data and results outside of the labs that currently drive the domain and increase secondary added-value of our experiments and efforts. We hope and expect that the initial guidelines offered here will evolve with the advancement of single-cell proteomics technologies[84], the increasing scale and sophistication of biological questions investigated by these technologies, and the integration with other data modalities, such as single-cell transcriptomics, spatial transcriptomics, imaging, electrophysiology, prioritized MS approaches, PTM-level and proteoform-level (i.e., top-down) single-cell proteomics methods. We invite the community to discuss these guidelines and contribute to their evolution. We hope to facilitate such broader contributions via an online portal at: single-cell.net/guidelines

## Acknowledgements
We thank the numerous contributors to these initial recommendations and the community as a whole for the body of work that supports our recommendations. We thank R. Gray Huffman for feedback and detailed edits. The guidelines in this article were formulated in large part during the workshops and through the discussions of the annual Single-Cell Proteomics Conference (https://single-cell.net).

## Competing interests
The authors declare no competing interests.